\documentclass{article}

\usepackage{amsmath}
\usepackage{amssymb}
\usepackage{algorithm}
\usepackage{algorithmic}

\usepackage{amsfonts}
\usepackage{graphics}
\usepackage{epsfig}

\newcommand{\reals}{\mathbb{R}}

\title{Exploiting Semidefinite Relaxations in Constraint Programming}

\author{Willem Jan van Hoeve\\
CWI, P.O. Box 94079, 1090 GB Amsterdam, The Netherlands\\
{\tt W.J.van.Hoeve@cwi.nl}}

\date{July 16, 2004}

\begin{document}

\maketitle 

\begin{abstract}
Constraint programming uses enumeration and search tree pruning to
solve combinatorial optimization problems. In order to speed up
this solution process, we investigate the use of semidefinite
relaxations within constraint programming.
In principle, we use the solution of a semidefinite relaxation to
guide the traversal of the search tree, using a limited
discrepancy search strategy. Furthermore, a semidefinite 
relaxation produces a bound for the solution value, which is used 
to prune parts of the search tree.
Experimental results on stable set and maximum clique problem 
instances show that constraint programming can indeed
greatly benefit from semidefinite relaxations.
\end{abstract}

\section{Introduction}
Constraint programming models for combinatorial optimization 
problems consist of variables on finite domains, constraints 
on those variables and an objective function to be optimized.
In general, constraint programming solvers use domain
value enumeration to solve combinatorial optimization problems. By
propagation of the constraints (i.e. removal of inconsistent values),
large parts of the resulting search tree may be pruned.
Because combinatorial optimization problems are NP-hard in general,
constraint propagation is essential to make constraint programming
solvers practically applicable. Another essential part concerns the
enumeration scheme, that defines and traverses a search tree. Variable
and value ordering heuristics as well as tree traversal heuristics
greatly influence the performance of the resulting constraint programming
solver.

In this work we investigate the possibility of using semidefinite
relaxations in constraint programming. This investigation involves 
the extraction of semidefinite relaxations from a constraint 
programming model, and the actual use of the relaxation inside the
solution scheme. We propose to use the solution of a 
semidefinite relaxation to define search tree ordering and traversal
heuristics. Effectively, this means that our enumeration scheme starts 
at the suggestion made by the semidefinite relaxation, and gradually 
scans a wider area around this solution. Moreover, we use the solution 
value of the semidefinite relaxation as a bound for the objective 
function, which results in stronger pruning. By applying a semidefinite
relaxation in this way, we hope to speed up the constraint programming
solver significantly. These ideas were motivated by a previous 
work~\cite{milano_hoeve02}, in which a linear relaxation was proved 
to be helpful in constraint programming.

We implemented our method and provide experimental results on the stable 
set problem and the maximum clique problem, two classical combinatorial 
optimization problems. We compare our method with a standard constraint
programming solver, and with specialized solvers for maximum clique 
problems. As computational results will show, our method obtains far
better results than a standard constraint programming solver. However, 
on maximum clique problems, the specialized solvers appear to be much 
faster than our method.

This paper is an extended and revised version of \cite{hoeve_cp03}.
In the current version, a more general view on the proposed method is 
presented. Namely, in \cite{hoeve_cp03} the method was proposed for
stable set problems only, while in this paper we propose the method 
to be applied to any constraint programming problem, although not 
all problems will be equally suitable. Furthermore, in 
\cite{hoeve_cp03} the method uses a subproblem generation framework
on which limited discrepancy search is applied. In the current work
it has been replaced by limited discrepancy search on single values, 
which is more concise while preserving the behaviour of the algorithm.
Finally, more experimental results are presented, including problem 
characterizations and instances of the DIMACS benchmark set for the 
maximum clique problem.

The outline of the paper is as follows. The next section gives a
motivation for the approach proposed in this work. Then, in
Section~\ref{sc:prel} some preliminaries on constraint and semidefinite
programming are given. A description of our solution framework is
given in Section~\ref{sc:framework}. In Section~\ref{sc:formulations}
we introduce the stable set problem and the maximum clique problem, 
integer optimization formulations and a semidefinite relaxation. 
Section~\ref{sc:results} presents the computational results. 
We conclude in Section~\ref{sc:conclusions} with a summary and 
future directions.

\section{Motivation}
NP-hard combinatorial optimization problems are often solved
with the use of a polynomially solvable relaxation. Often (continuous) 
linear relaxations are chosen for this purpose. Also within 
constraint programming, linear relaxations are widely used, see 
\cite{focacci04} for an overview. 
Let us first motivate why in this paper a semidefinite relaxation is 
used rather than a linear relaxation. For some problems, for instance
for the stable set problem, standard linear relaxations are not very 
tight and not informative. 
One way to overcome this problem is to identify and 
add linear constraints that strengthen the relaxation. But it may be 
time-consuming to identify such constraints, and by enlarging the 
model the solution process may slow down.

On the other hand, several papers on approximation theory 
following~\cite{GW_maxcut} have shown the tightness of semidefinite 
relaxations. However, being tighter, semidefinite programs are more 
time-consuming to solve than linear programs in practice. Hence one has 
to trade strength for computation time. For some (large scale) 
applications, semidefinite relaxations are well-suited to be used within 
a branch and bound framework (see for instance \cite{karisch2000}).
Moreover, our intention is not to solve a relaxation at every node of 
the search tree. Instead, we propose to solve only once a relaxation, 
before entering the search tree. Therefore, we are willing to make the 
trade-off in favour of the semidefinite relaxation.

Finally, investigating the possibility of using semidefinite 
relaxations in constraint programming is worthwile in itself.
To our knowledge the cross-ferti\-li\-zation of semidefinite 
programming and constraint programming has not yet been investigated.
Hence, this paper should be seen as a first step toward the cooperation 
of constraint programming and semidefinite programming.

\section{Preliminaries}\label{sc:prel}

\subsection{Constraint Programming}\label{ssc:cp}
In this section we briefly introduce the basic concepts of constraint
programming that are used in this paper.
A thorough explanation of the principles of constraint programming can
be found in \cite{apt2003}. 

A constraint programming model consists of a set of variables, 
corresponding variable domains, and a set of constraints
restricting those variables. In case of optimization problems, also an
objective function is added. In this work the
variable domains are assumed to be finite. A constraint $c$ is
defined as a subset of the Cartesian product of the domains of the
variables that are in $c$. Constraints may be of any form (linear,
nonlinear, logical, symbolic, etcetera), provided that the constraint
programming solver contains an algorithm to check its satisfiability,
or even to identify globally inconsistent domain values.

Basically, a constraint programming solver tries to find a solution
of the model by enumerating all possible variable-value assignments
such that the constraints are all satisfied. Because there are
exponentially many possible assignments, constraint propagation is
needed to prune large parts of the corresponding search
tree. Constraint propagation tries to remove inconsistent values from
variable domains before the variables are actually
instantiated. Hence, one doesn't need to generate the whole search
tree, but only a part of it, while still preserving a complete (exact)
solution scheme. The general solution scheme is an iterative
process in which branching decisions are made, and the effects are 
propagated subsequently.

Variable and value ordering heuristics, which define the search tree, 
greatly influence the constraint propagation, and with that the
performance of the solver. If no suitable variable and value ordering 
heuristics are available, constraint programming solvers often
use a lexicographic variable and value ordering, and depth-first 
search to traverse the tree. However, when good heuristics are
available, they should be applied.

When `perfect' value and variable ordering heuristics are followed, 
it will lead us directly to the optimal solution (possibly unproven). 
Although perfect heuristics are often not available, some heuristics 
come pretty close. In such cases, one should try to deviate from the 
first heuristic solution as little as possible. This is done by 
traversing the search tree using a limited discrepancy search strategy
(LDS) \cite{harvey95} instead of depth-first search.

LDS is organized in waves of increasing discrepancy from the first 
solution provided by the heuristic. The first wave (discrepancy 0) 
exactly follows the heuristic. The next waves 
(discrepancy $i$, with $i>0$), explore all the solutions that can 
be reached when $i$ derivations from the heuristic are made. 
Typically, LDS is applied until a maximum discrepancy has been 
reached, say 3 or 4. Although being incomplete (inexact), the 
resulting strategy often finds good solutions very fast, provided 
that the heuristic is informative. Of course LDS can also be applied 
until all possible discrepancies have been considered, resulting
in a complete strategy.

\subsection{Semidefinite Programming}\label{ssc:sdp}
In this section we briefly introduce semidefinite programming. 
A large number of references to papers concerning semidefinite
programming are on the web pages \cite{sdp_helmberg} and
\cite{sdp_alizadeh}. A general introduction on semidefinite
programming applied to combinatorial optimization is given in
\cite{goemans_rendl} and \cite{laurent_rendl}.

Semidefinite programming makes use of positive semidefinite matrices
of variables. A matrix $X \in \reals^{n \times n}$ is said to be
positive semidefinite (denoted by $X \succeq 0$) when $y^{\sf T}Xy
\geq 0$ for all vectors $y \in \reals^n$. Semidefinite programs have
the form 
\begin{equation}\label{eq:sdp_general}
\begin{array}{rll}
{\rm max}  & {\rm tr} (WX) \\
{\rm s.t.} & {\rm tr} (A_j X) \leq b_j & (j=1,\dots,m)\\
           & X \succeq 0.
\end{array}
\end{equation}
Here ${\rm tr}(X)$ denotes the trace of $X$, which is the sum of
its diagonal elements, i.e. ${\rm tr}(X) = \sum_{i=1}^{n} X_{ii}$. 
The matrix $X$, the cost matrix $W \in \reals^{n \times n}$ and the 
constraint matrices $A_j \in \reals^{n \times n}$ are supposed to be 
symmetric. The $m$ reals $b_j$ and the $m$ matrices $A_j$ define 
$m$ constraints.

We can view semidefinite programming as an extension of linear
programming. Namely, when the matrices $W$ and $A_j \;(j=1, \dots, m)$
are all supposed to be diagonal matrices\footnote{A diagonal matrix is
a matrix with nonnegative values on its diagonal entries only.}, the 
resulting semidefinite program is equal to a linear program, where the 
matrix $X$ is replaced by a non-negative vector of variables 
$x \in \reals^n$. In particular, then a semidefinite programming 
constraint tr$(A_j X) \leq b_j$ corresponds to a linear programming 
constraint $a_j^{\sf T}x \leq b_j$, where $a_j$ represents the diagonal 
of $A_j$.

Theoretically, semidefinite programs have been proved to be
polynomially solvable to any fixed precision using the so-called
ellipsoid method (see for instance~\cite{GLS88}). In practice,
nowadays fast `interior point' methods are being used for this purpose
(see~\cite{alizadeh95} for an overview). 

Semidefinite programs may serve as a continuous relaxation for
(integer) combinatorial optimization problems. Unfortunately, it is
not a trivial task to obtain a computationally efficient semidefinite
program that provides a tight solution for a given problem. However,
for a number of combinatorial optimization problems such semidefinite
relaxations do exist, for instance the stable set problem, the maximum
cut problem, quadratic programming problems, the maximum
satisfiability problem, and many others (see \cite{laurent_rendl} for
an overview).

\section{Solution Framework} \label{sc:framework}
The skeleton of our solution framework is formed by the constraint 
programming enumeration scheme, or search tree, as explained in 
Section~\ref{ssc:cp}. Within this skeleton, we want to use the solution 
of a semidefinite relaxation to define the variable and value ordering
heuristics. In this section we first show how to extract a
semidefinite relaxation from a constraint programming model. Then we 
give a description of the usage of the relaxation within the 
enumeration scheme.

\subsection{Building a Semidefinite Relaxation}
We start from a constraint programming model consisting of a set of
variables $\{v_1, \dots, v_n\}$, corresponding finite domains $\{D_1,
\dots, D_n\}$, a set of constraints and an objective 
function. From this model we need to extract a semidefinite
relaxation. In general, a relaxation is obtained by removing or 
replacing one or more constraints such that all solutions are preserved.
If it is possible to identify a subset of constraints for which a 
semidefinite relaxation is known, this relaxation can be used inside 
our framework. Otherwise, we need to build up a relaxation from scratch. 
This can be done in the following way.

If all domains $D_1, \dots, D_n$ are binary, a semidefinite relaxation 
can be extracted using a method proposed by~\cite{laurent97connections}, 
which is explained below. In general,
however, the domains are non-binary. In that case, we transform the
variables $v_i$ and the domains $D_i$ into corresponding binary
variables $x_{ij}$ for $i \in \{1, \dots, n\}$ and $j \in D_i$:
\begin{equation}\label{eq:transform}
\begin{array}{c}
v_i = j \Leftrightarrow x_{ij} = 1,\\
v_i \neq j \Leftrightarrow x_{ij} = 0.
\end{array}
\end{equation}
We will then use the binary variables $x_{ij}$ to construct a semidefinite
relaxation. Of course, the transformation has consequences for the
constraints also, which will be discussed below.

The method to transform a model with binary variables into a
semidefinite relaxation, presented in \cite{laurent97connections},
is the following.
Let $d \in \{0,1\}^N$ be a vector of binary variables, where $N$ is
a positive integer. Construct the $(N+1) \times (N+1)$ variable matrix 
$X$ as 
\begin{displaymath}
X = \binom{1}{d} (1 \; d^{\sf T}) = 
\left(
\begin{array}{cc}
1 & d^{\sf T} \\
d &  d d^{\sf T}
\end{array}\right).
\end{displaymath}
Then $X$ can be constrained to satisfy 
\begin{eqnarray}
X \succeq 0 \label{eq:psd} \\
X_{ii} = X_{0i} & \forall i \in \{1, \dots, N\} \label{eq:diag}
\end{eqnarray}
where the rows and columns of $X$ are indexed from 0 to $N$. Condition
(\ref{eq:diag}) expresses the fact that $d_i^2 = d_i$, which is
equivalent to $d_i \in \{0,1\}$. Note however that the latter
constraint is relaxed by requiring $X$ to be positive
semidefinite.

The matrix $X$ contains the variables to model our semidefinite
relaxation. Obviously, the diagonal entries (as well as the first row
and column) of this matrix represent the binary variables from which
we started. Using these variables, we need to rewrite (a part of) the
original constraints into the form of program~(\ref{eq:sdp_general})
in order to build the semidefinite relaxation.

In case the binary variables are obtained from transformation 
(\ref{eq:transform}), not all constraints may be trivially transformed
accordingly. Especially because the original constraints may be of any
form. The same holds for the objective function. On the other hand, as
we are constructing a relaxation, we may choose among the set of
constraints an appropriate subset to include in the
relaxation. Moreover, the constraints itself are allowed to be
relaxed. Although there is no `recipe' to transform any given original
constraint into the form of program~(\ref{eq:sdp_general}), one may 
use results from the literature \cite{laurent_rendl}. For instance, for 
linear constraints on binary variables a straightforward translation is 
given in Section~\ref{ssc:sdp}.

\subsection{Applying the Semidefinite Relaxation}
\begin{figure}
\begin{center}
\epsfig{figure=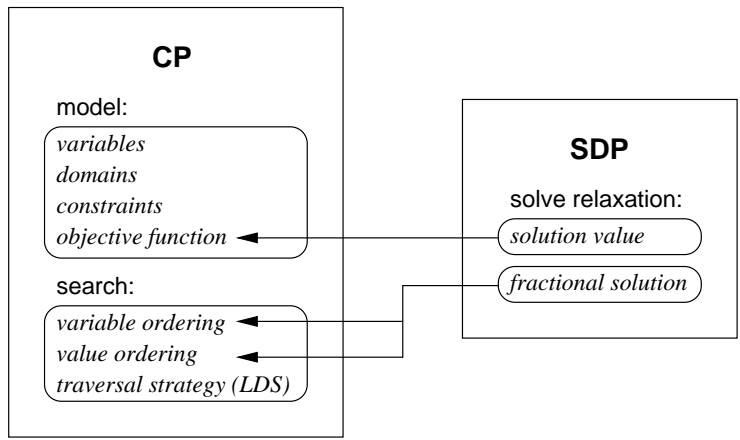}
\end{center}
\caption{Communication between constraint programming (CP) and semidefinite programming (SDP).}\label{fg:method}
\end{figure}

At this point, we have either identified a subset of constraints for 
which a semidefinite relaxation exists, or built up our own relaxation.
Now we show how to apply the solution to the semidefinite relaxation
inside the constraint programming framework, also depicted in 
Figure~\ref{fg:method}.
In general, the solution to the semidefinite relaxation yields 
fractional values for its variable matrix. For example, the diagonal 
variables $X_{ii}$ of the above matrix will be assigned to a value 
between 0 and 1. These fractional values serve as an indication for the 
original constraint programming variables. Consider for example again the
above matrix $X$, and suppose it is obtained from non-binary original 
variables, by transformation (\ref{eq:transform}). Assume that variable
$X_{ii}$ corresponds to the binary variable $x_{jk}$ (for some integer 
$j$ and $k$), which corresponds to $v_j = k$, where $v_j$ is a constraint 
programming variable and $k \in D_j$. If variable $X_{ii}$ is close to 1, 
then also $x_{jk}$ is supposed to be close to 1, which corresponds to 
assigning $v_j = k$.

Hence, our variable and value ordering heuristics for the constraint 
programming variables are based upon the fractional solution 
values of the corresponding variables in the semidefinite relaxation. 
Our variable ordering heuristic is to select first the constraint 
programming variable for which the corresponding fractional solution 
is closest to the corresponding integer solution. Our value ordering 
heuristic is to select first the corresponding suggested 
value. For example, consider again the above matrix $X$, obtained from
non-binary variables by transformation (\ref{eq:transform}). 
We select first the variable $v_j$ for which $X_{ii}$, representing the 
binary variable $x_{jk}$, is closest to 1, for some $k \in D_j$ and 
corresponding integer $i$. Then we assign value $k$ to variable $v_j$.
We have also implemented a randomized variant of the above variable 
ordering heuristic. In the randomized case, the selected variable is
accepted with a probability proportional to the corresponding 
fractional value.

We expect the semidefinite relaxation to provide promising values.
Therefore the resulting search tree will be traversed using 
limited discrepancy search, defined in Section~\ref{ssc:cp}. 
A last remark concerns the solution value of the semidefinite 
relaxation, which is used as a bound on the objective function in 
the constraint programming model. If this bound is tight, which is
the case in our experiments, it leads to more propagation and a 
smaller search space.

\section{The Stable Set Problem and Maximum Clique Problem} \label{sc:formulations}
This section describes the stable set problem and the maximum clique 
problem (see \cite{pardalos_xue,bomze} for a survey), on which we have 
tested our algorithm. First we give their 
definitions, and the equivalence of the two problems. Then we will focus
on the stable set problem, and formulate it as an integer optimization
problem. From this, a semidefinite relaxation is inferred.

\subsection{Definitions}
Consider an undirected weighted graph $G=(V,E)$, where $V = 
\{1, \dots, n\}$ is the set of vertices and $E$ a subset of edges 
$\{(i,j)| i,j \in V, i \neq j \}$ of $G$, with $|E|=m$. To each 
vertex $i \in V$ a weight $w_i \in \reals$ is assigned (without 
loss of generality, we can assume all weights to be nonnegative). 

A stable set is a set $S \subseteq V$ such that no two vertices 
in $S$ are joined by an edge in $E$.
The stable set problem is the problem of finding a stable set of
maximum total weight in $G$. This value is called the stable set
  number of $G$ and is denoted by $\alpha(G)$\footnote{In the literature
$\alpha(G)$ usually denotes the unweighted stable set number. The weighted
stable set number is then denoted as $\alpha_w(G)$. In this work, it is
not necessary to make this distinction.}. In the unweighted case (when all 
weights are equal to 1), this problem amounts to the maximum cardinality 
stable set problem, which has been shown to be already 
NP-hard~\cite{papadimitriou_steiglitz}.

A clique is a set $C \subseteq V$ such that every two vertices 
in $C$ are joined by an edge in $E$. The maximum clique problem
is the problem of finding a clique of maximum total weight in $G$. 
This value is called the clique number of $G$ and is denoted by 
$\omega(G)$\footnote{$\omega_w(G)$ is defined similar to $\alpha_w(G)$
and also not distinguished in this paper.}.

The complement graph of $G$ is $\overline{G} = (V,\overline{E})$, with 
the same set of vertices $V = \{1, \dots, n\}$, but with edge set
$\overline{E} = \{(i,j)| i,j \in V, (i,j) \notin E, i \neq j \}$. It is 
well known that $\alpha(G) = \omega(\overline{G})$. Hence, a maximum 
clique problem can be translated into a stable set problem on the 
complement graph. We will do exactly this in our implementation, and 
focus on the stable set problem, for which good semidefinite relaxations
exist.

\subsection{Integer Optimization Formulation}\label{ssc:int_form}
Let us first consider an integer linear programming formulation
for the stable set problem.
We introduce binary variables to indicate whether or not a vertex
belongs to the stable set $S$. So, for $n$ vertices, we have $n$
integer variables $x_i$ indexed by $i \in V$, with initial
domains $\{0,1\}$. In this way, $x_i=1$ if vertex $i$ is in 
$S$, and $x_i = 0$ otherwise. We can now state the objective
function, being the sum of the weights of vertices that are in
$S$, as $\sum_{i=1}^n w_i x_i$. Finally, we define the
constraints that forbid two adjacent vertices to be both inside $S$
as $x_i + x_j \leq 1$, for all edges $(i,j) \in E$. Hence the
integer linear programming model becomes:
\begin{equation} \label{eq:ilp_form}
\begin{array}{rrrl}
\alpha(G) = & {\rm max} & \sum_{i=1}^{n} w_i x_i \\
&  {\rm s.t.} & x_i + x_j \leq 1 & \forall (i,j) \in E \\
&   & x_i \in \{0,1\}  & \forall i \in V.
\end{array}
\end{equation}

Another way of describing the same solution set is presented by the
following integer quadratic program
\begin{equation}\label{eq:quadratic}
\begin{array}{rrrl}
\alpha(G) = & {\rm max} & \sum_{i=1}^{n} w_i x_i \\
 & {\rm s.t.}& x_i x_j = 0 & \forall (i,j) \in E \\
 & & x_i^2 = x_i & \forall i \in V.
\end{array}
\end{equation}
Note that here the constraint $x_i \in \{0,1\}$ is replaced by $x_i^2
= x_i$, similar to condition~(\ref{eq:diag}) in Section~\ref{sc:framework}. 
This quadratic formulation will be used below to infer a semidefinite 
relaxation of the stable set problem.

In fact, both model (\ref{eq:ilp_form}) and model (\ref{eq:quadratic})
can be used as a constraint programming model. We have chosen the first
model, since the quadratic constraints take more time to propagate
than the linear constraints, while having the same pruning power.
To infer the semidefinite relaxation, however, we will use the 
equivalent model (\ref{eq:quadratic}).

\subsection{Semidefinite Programming Relaxation}
The integer quadratic program (\ref{eq:quadratic}) gives rise to a
well-known semidefinite relaxation introduced by Lov\'asz \cite{lovasz79} 
(see~\cite{GLS88} for a comprehensive treatment). The
value of the objective function of this relaxation has been named the
theta number of a graph $G$, indicated by $\vartheta(G)$.
For its derivation into a form similar to 
program~(\ref{eq:sdp_general}), we will follow the same idea as in 
Section~\ref{sc:framework} for the general case.

As our constraint programming model uses binary variables already, we 
can immediately define the $(n+1) \times (n+1)$ matrix variable $X$ of 
our relaxation as 
\begin{displaymath}
X = \left(
\begin{array}{cc}
1 & x^{\sf T} \\
x &  x x^{\sf T} \\
\end{array}\right)
\end{displaymath}
where the binary vector $x$ again represents the stable set, as in 
Section~\ref{ssc:int_form}. First we impose the constraints
\begin{eqnarray}
X \succeq 0 \\
X_{ii} = X_{0i} & \forall i \in \{1, \dots, n\}
\end{eqnarray}
as described in Section~\ref{sc:framework}. 
Next we translate the edge constraints $x_i x_j = 0$ from 
program~(\ref{eq:quadratic}) into $X_{ij} = 0$, because $X_{ij}$ 
represents $x_i x_j$. In order to translate the objective function, 
we first define the $(n+1) \times (n+1)$ weight matrix $W$ as 
\begin{displaymath}
\begin{array}{ll}
W_{ii} = w_i & \forall i \in \{1, \dots, n\}, \\
W_{ij} = 0   & \forall i \in \{0, \dots, n\}, j \in \{0, \dots, n\}, 
i \neq j.
\end{array}
\end{displaymath}
Then the objective function translates into ${\rm tr(W X)}$.
The semidefinite relaxation thus becomes
\begin{equation} \label{eq:theta1}
\begin{array}{rrcl}
\vartheta(G) = & {\rm max} & {\rm tr(W X)} \\
& {\rm s.t.}& X_{ii} = X_{0i} & \forall i \in V \\
& & X_{ij} = 0 & \forall (i,j) \in E \\
& & X \succeq 0.
\end{array}
\end{equation}

Note that program~(\ref{eq:theta1}) can easily be rewritten into the
general form of program~(\ref{eq:sdp_general}). Namely, $X_{ii} =
X_{0i}$ is equal to tr$(A_i X) = 0$ where the $(n+1) \times (n+1)$ 
matrix $A_i$ consists of all zeroes, except for $(A_i)_{ii} = 1$, 
$(A_i)_{i0} = -\frac{1}{2}$ and $(A_i)_{0i} = -\frac{1}{2}$,
which makes the corresponding right-hand side ($b_i$ entry) equal to 0 
(similarly for the edge constraints).

The theta number also arises from other formulations, different from
the above, see \cite{GLS88}. In our implementation we have used the
formulation that has been shown to be computationally most efficient
among those alternatives \cite{gruber_rendl}. Let us introduce that
particular formulation (called $\vartheta_3$ in \cite{GLS88}).
Again, let $x \in \{0,1\}^n$ be a vector of
binary variables representing a stable set. Define the $n \times n$
matrix $\tilde{X} = \xi \xi^{\sf T}$ where
$\xi_i = \frac{\sqrt{w_i}}{\sqrt{\sum_{j=1}^n w_j x_j}} x_i$. Furthermore,
let the $n\times n$ cost matrix $U$ be defined as $U_{ij} = \sqrt{w_i  w_j}$
for $i,j \in V$. Observe that in these definitions we exploit the fact
that $w_i \geq 0$ for all $i \in V$. The following semidefinite program
\begin{equation} \label{eq:theta2}
\begin{array}{rrrl}
\vartheta(G) = & {\rm max} & {\rm tr}(U\tilde{X}) \\
& {\rm s.t.}& {\rm tr} (\tilde{X}) = 1 \\
& & \tilde{X}_{ij} = 0 & \forall (i,j) \in E \\
& & \tilde{X} \succeq 0
\end{array}
\end{equation}
has been shown to also give the theta number of $G$, see \cite{GLS88}. 
When (\ref{eq:theta2}) is solved to optimality, the scaled diagonal 
element $\vartheta(G) \tilde{X}_{ii}$ (a fractional value between 0 
and 1) serves as an indication for the value of $x_i$ ($i \in V$)
in a maximum stable set (see for instance \cite{gruber_rendl}).
Again, it is not difficult to rewrite program (\ref{eq:theta2}) into
the general form of program~(\ref{eq:sdp_general}).

Program~(\ref{eq:theta2}) uses matrices of dimension $n$ and $m+1$
constraints, while program~(\ref{eq:theta1}) uses matrices of
dimension $n+1$ and $m+n$ constraints. This gives an indication why
program~(\ref{eq:theta2}) is computationally more efficient.

\section{Computational Results} \label{sc:results}
All our experiments are performed on a Sun Enterprise 450 
(4 X UltraSPARC-II 400MHz) with maximum 2048 Mb memory size, on which 
our algorithms only use one processor of 400MHz at a time.
As constraint programming solver we use the ILOG Solver 
library, version 5.1 \cite{ilog51}. As semidefinite programming solver, 
we use CSDP version 4.1 \cite{csdp}, with the optimized ATLAS 3.4.1 
\cite{atlas} and LAPACK 3.0 \cite{lapack} libraries for matrix computations.
The reason for our choices is that both solvers are among the fastest in 
their field, and because ILOG Solver is written in C++, and CSDP is 
written in C, they can be hooked together relatively easy.

We distinguish two algorithms to perform our experiments. The first 
algorithm is a sole constraint programming solver, which uses a standard 
enumeration strategy. This means we use a lexicographic variable ordering, 
and we select domain value 1 before value 0. 
The resulting search tree is traversed using a depth-first search strategy. 
After each branching decision, its effect is directly propagated through 
the constraints. As constraint programming model we have used 
model (\ref{eq:ilp_form}), as was argued in Section~\ref{ssc:int_form}.

The second algorithm is the one proposed in Section~\ref{sc:framework}. 
It first solves the semidefinite program (\ref{eq:theta2}), and then 
calls the constraint programming solver. In this case, we use the
randomized variable ordering heuristic, defined by the solution of the 
semidefinite relaxation. The resulting search tree is traversed using a 
limited discrepancy search strategy. In fact, in order to improve our 
starting solution, we repeat the search for the first solution $n$ 
times, (where $n$ is the number of variables), and the best solution 
found is the heuristic solution to be followed by the limited discrepancy 
search strategy.

\subsection{Characterization of Problem Instances}\label{ssc:properties}
We will first identify general characteristics of the constraint 
programming solver and the semidefinite programming solver applied to
stable set problems. It appears that both solvers 
are highly dependent on the edge density of the graph, i.e. 
$\frac{m}{\frac{1}{2}(n^2 - n)}$ for a graph with $n$ vertices and
$m$ edges. We therefore 
generated random graphs on 30, 40, 50 and 60 vertices,
with density ranging from 0.01 up to 0.95. Our aim is to identify the
hardness of the instances for both solvers, parametrized by the density.
Based upon this information, we can identify the kind of problems 
our algorithm is suitable for. 

\begin{figure}
\begin{center}
\epsfig{figure=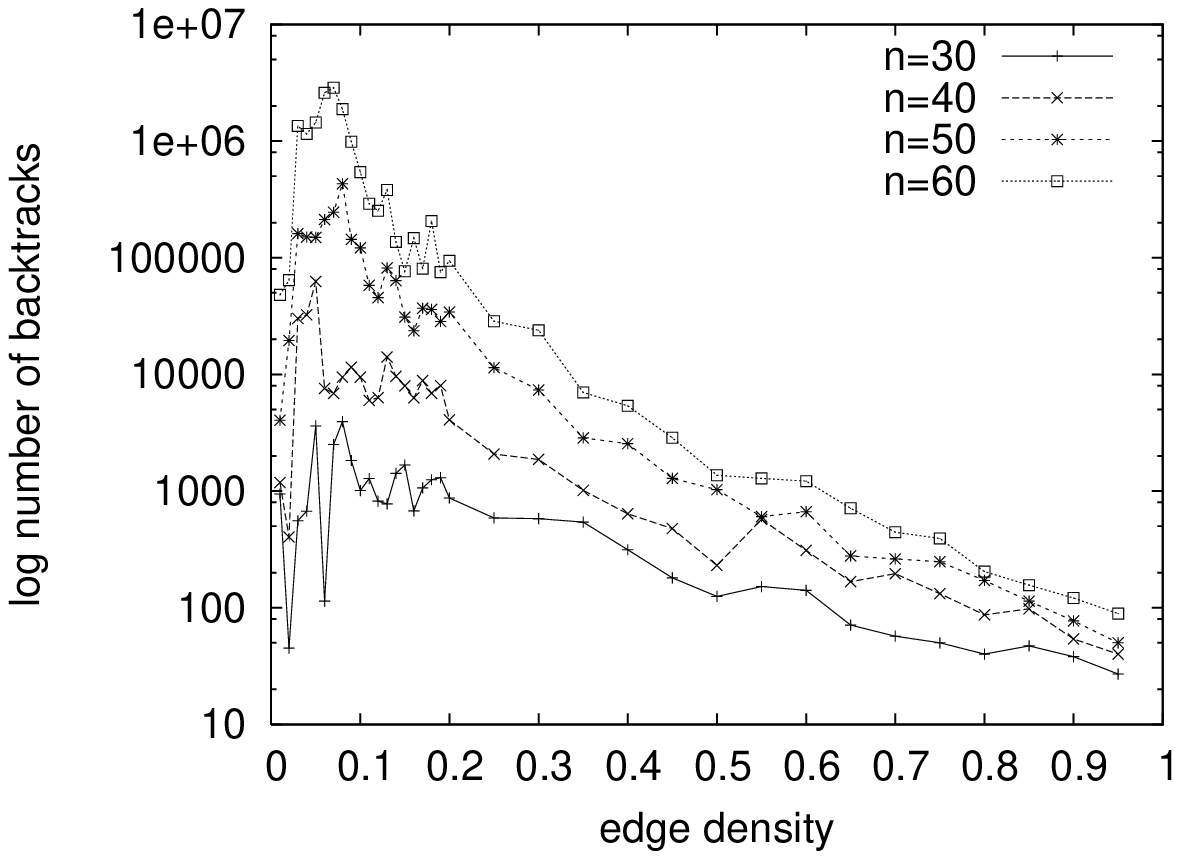,width=.49\textwidth} \hfill
\epsfig{figure=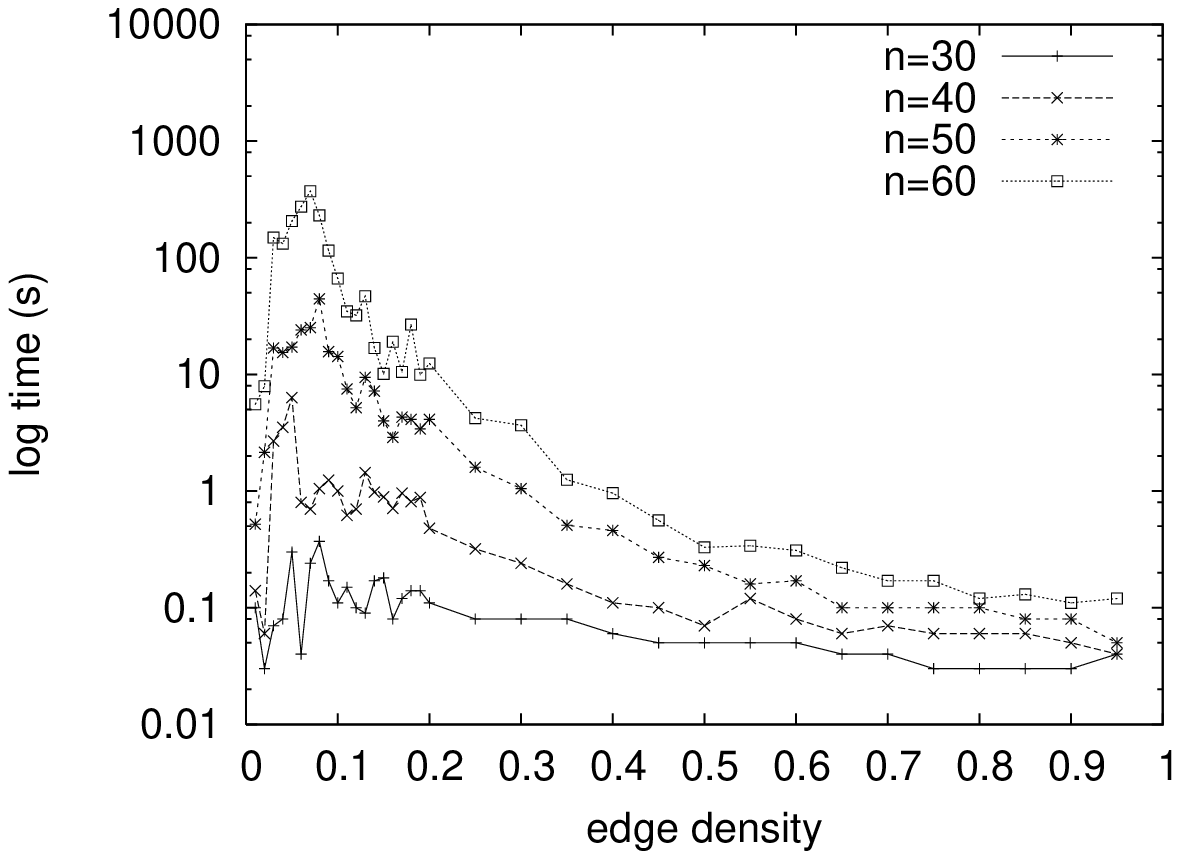,width=.49\textwidth}
\end{center}
\caption{Performance of the constraint programming solver on random 
instances with $n$ vertices.} \label{fg:perf_cp}
\end{figure}
\begin{figure}
\begin{center}
\epsfig{figure=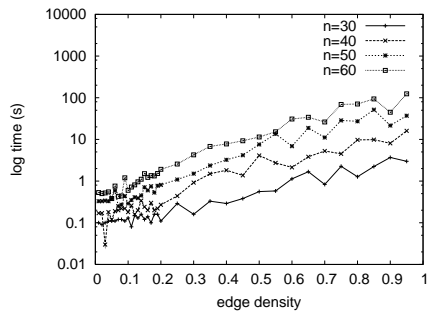,width=.49\textwidth}
\end{center}
\caption{Performance of the semidefinite programming solver on random 
instances with $n$ vertices.} \label{fg:perf_sdp}
\end{figure}

We have plotted the performance of both solvers in Figure~\ref{fg:perf_cp} 
and Figure~\ref{fg:perf_sdp}. Here the constraint programming solver
solves the problems to optimality, while the semidefinite programming 
solver only solves the semidefinite relaxation.
For the constraint programming solver, we depict both the number of 
backtracks and the time needed to prove optimality. 
For the semidefinite programming solver we only 
plotted the time needed to solve the relaxation. Namely, this solver
does not use a tree search, but a so-called primal-dual interior point
algorithm. Note that we use a log-scale for time and number of backtracks
in these pictures.

From these figures, we can conclude that the constraint programming solver
has the most difficulties with instances up to density around 0.2.
Here we see the effect of constraint propagation. As the 
number of constraints increases, the search tree can be heavily pruned.
On the other hand, our semidefinite relaxation suffers from every edge that
is added. As the density increases, the semidefinite program increases
accordingly, as well as its computation time. Fortunately, for the 
instances up to 0.2, the computation time for the semidefinite relaxation 
is very small. Consequently, our algorithm is expected to behave best 
for graphs that have edge density up to around 0.2. For graphs with a 
higher density, the constraint programming solver is expected to use
less time than the semidefinite programming solver, which makes the 
application of our method unnecessary.

\subsection{Random Weighted and Unweighted Graphs}
Our first experiments are performed on random weighted and unweighted
graphs. We generated graphs with 50, 75, 100, 125 and 150 vertices 
and edge density from 0.05, 010 and 0.15, corresponding to the interesting
problem area. The results are presented in Table~\ref{tb:random}.
Unweighted graphs on $n$ vertices and edge density $r$ are named 
`g$n$d$r$'. Weighted graphs are similarly named `wg$n$d$r$'.

\begin{table}[t]
\caption{Computational results on random graphs, with $n$ vertices and $m$ 
edges. All times are in seconds. The time limit is set to 1000 seconds.} 
\label{tb:random}
\begin{center}
\renewcommand{\arraystretch}{1.1}
\setlength{\tabcolsep}{0.09cm}
\tiny
\setlength{\doublerulesep}{\arrayrulewidth} 
\begin{tabular}{lrrcrcrrrcrcrrr} \\ \hline\hline
\multicolumn{5}{c}{\bf instance} & & \multicolumn{3}{c}{\bf CP} & & \multicolumn{5}{c}{\bf SDP + CP} \\ 
\cline{1-5} \cline{7-9} \cline{11-15}
 & & & edge &  & & & total & back- &  &  & best & sdp & total & back- \\
name & $n$ & $m$ & density & $\alpha$ & & $\alpha$ & time & tracks & & $\alpha$ & discr & time & time & tracks \\ \hline
g50d005 & 50 & 70 & 0.06 & 27 &  & 27 & 5.51 & 50567 &  & 27 & 0 & 0.26 & 0.27 & 0 \\
g50d010 & 50 & 114 & 0.09 & 22 &  & 22 & 28.54 & 256932 &  & 22 & 0 & 0.35 & 0.36 & 0 \\
g50d015 & 50 & 190 & 0.16 & 17 &  & 17 & 5.83 & 48969 &  & 17 & 0 & 0.49 & 0.49 & 0 \\
g75d005 & 75 & 138 & 0.05 & 36 &  & $\geq$ 35 & limit &  &  & 36 & 0 & 0.72 & 0.73 & 0 \\
g75d010 & 75 & 282 & 0.10 & $\geq$ 25 &  & $\geq$ 25 & limit &  &  & $\geq$ 25 & 5 & 1.4 & limit &  \\
g75d015 & 75 & 426 & 0.15 & 21 &  & 21 & 170.56 & 1209019 &  & 21 & 0 & 2.81 & 664.92 & 1641692 \\
g100d005 & 100 & 254 & 0.05 & 43 &  & $\geq$ 40 & limit &  &  & 43 & 0 & 2.07 & 2.1 & 0 \\
g100d010 & 100 & 508 & 0.10 & $\geq$ 31 &  & $\geq$ 30 & limit &  &  & $\geq$ 31 & 0 & 4.94 & limit &  \\
g100d015 & 100 & 736 & 0.15 & $\geq$ 24 &  & $\geq$ 24 & limit &  &  & $\geq$ 24 & 4 & 9.81 & limit &  \\
g125d005 & 125 & 393 & 0.05 & $\geq$ 49 &  & $\geq$ 44 & limit &  &  & $\geq$ 49 & 1 & 4.92 & limit &  \\
g125d010 & 125 & 791 & 0.10 & $\geq$ 33 &  & $\geq$ 30 & limit &  &  & $\geq$ 33 & 6 & 12.58 & limit &  \\
g125d015 & 125 & 1160 & 0.15 & $\geq$ 27 &  & $\geq$ 24 & limit &  &  & $\geq$ 27 & 1 & 29.29 & limit &  \\
g150d005 & 150 & 545 & 0.05 & $\geq$ 52 &  & $\geq$ 44 & limit &  &  & $\geq$ 52 & 3 & 10.09 & limit &  \\
g150d010 & 150 & 1111 & 0.10 & $\geq$ 38 &  & $\geq$ 32 & limit &  &  & $\geq$ 38 & 4 & 27.48 & limit &  \\
g150d015 & 150 & 1566 & 0.14 & $\geq$ 29 &  & $\geq$ 26 & limit &  &  & $\geq$ 29 & 8 & 57.86 & limit &  \\
wg50d005 & 50 & 70 & 0.06 & 740 &  & 740 & 4.41 & 30528 &  & 740 & 0 & 0.29 & 0.3 & 0 \\
wg50d010 & 50 & 126 & 0.10 & 636 &  & 636 & 3.12 & 19608 &  & 636 & 0 & 0.41 & 0.41 & 0 \\
wg50d015 & 50 & 171 & 0.14 & 568 &  & 568 & 4.09 & 25533 &  & 568 & 0 & 0.59 & 4.93 & 13042 \\
wg75d005 & 75 & 128 & 0.05 & 1761 &  & 1761 & 744.29 & 4036453 &  & 1761 & 0 & 1.05 & 1.07 & 0 \\
wg75d010 & 75 & 284 & 0.10 & 1198 &  & 1198 & 325.92 & 1764478 &  & 1198 & 13 & 1.9 & 924.2 & 1974913 \\
wg75d015 & 75 & 409 & 0.15 & 972 &  & 972 & 40.31 & 208146 &  & 972 & 0 & 3.62 & 51.08 & 87490 \\
wg100d005 & 100 & 233 & 0.05 & 2302 &  & $\geq$ 2176 & limit &  &  & 2302 & 0 & 2.59 & 2.62 & 0 \\
wg100d010 & 100 & 488 & 0.10 & $\geq$ 1778 &  & $\geq$ 1778 & limit &  &  & $\geq$ 1778 & 2 & 6.4 & limit &  \\
wg100d015 & 100 & 750 & 0.15 & $\geq$ 1412 &  & $\geq$ 1412 & limit &  &  & $\geq$ 1412 & 2 & 15.21 & limit &  \\
wg125d005 & 125 & 372 & 0.05 & $\geq$ 3779 &  & $\geq$ 3390 & limit &  &  & $\geq$ 3779 & 3 & 5.39 & limit &  \\
wg125d010 & 125 & 767 & 0.10 & $\geq$ 2796 &  & $\geq$ 2175 & limit &  &  & $\geq$ 2796 & 0 & 18.5 & limit &  \\
wg125d015 & 125 & 1144 & 0.15 & $\geq$ 1991 &  & $\geq$ 1899 & limit &  &  & $\geq$ 1991 & 4 & 38.24 & limit &  \\
wg150d005 & 150 & 588 & 0.05 & $\geq$ 4381 &  & $\geq$ 3759 & limit &  &  & $\geq$ 4381 & 3 & 13.57 & limit &  \\
wg150d010 & 150 & 1167 & 0.10 & $\geq$ 3265 &  & $\geq$ 2533 & limit &  &  & $\geq$ 3265 & 9 & 40.68 & limit &  \\
wg150d015 & 150 & 1630 & 0.15 & $\geq$ 2828 &  & $\geq$ 2518 & limit &  &  & $\geq$ 2828 & 11 & 82.34 & limit &  \\ \hline\hline
\\
\end{tabular}
\normalsize
\end{center}
\end{table}

The first five columns of the table are dedicated to the instance, 
reporting its name, the number of vertices $n$ and edges $m$, the edge 
density and the (best known) value of a stable set, $\alpha$. The
next three columns (CP) present the performance of the constraint 
programming solver, reporting the best found estimate of $\alpha$,
the total time and the total number of backtracks needed to prove 
optimality. The last five columns (SDP+CP) present the performance
of our method, where also a column has been added for the discrepancy 
of the best found value (best discr), and a column for the time needed
by the semidefinite programming solver (sdp time).

Table~\ref{tb:random} shows that our approach always finds a better 
(or equally good) estimate for $\alpha$ than the standard constraint 
programming approach. This becomes more obvious for larger $n$.
However, there are two (out of 30) instances in which our method needs 
substantially more time to achieve this result (g75d015 and wg75d010).
A final observation concerns the discrepancy of the best found 
solutions. Our method appears to find those (often optimal) solutions
at rather low discrepancies.

\subsection{Graphs Arising from Coding Theory}
The next experiments are performed on structured (unweighted) graphs 
arising from coding theory, obtained from \cite{sloane}. We have used 
those instances that were solvable in reasonable time by the 
semidefinite programming solver (here reasonable means within 1000 
seconds). For these instances, the value of $\alpha$ happened to be 
known already.

\begin{table}
\caption{Computational results on graphs arising from coding theory, with
$n$ vertices and $m$ edges. All times are in seconds. The time limit is 
set to 1000 seconds.} 
\label{tb:sloane}
\begin{center}
\renewcommand{\arraystretch}{1.1}
\setlength{\tabcolsep}{0.12cm}
\tiny
\setlength{\doublerulesep}{\arrayrulewidth} 
\begin{tabular}{lrrcrcrrrcrcrrr} \\ \hline\hline
\multicolumn{5}{c}{\bf instance} & & \multicolumn{3}{c}{\bf CP} & & \multicolumn{5}{c}{\bf SDP + CP} \\ 
\cline{1-5} \cline{7-9} \cline{11-15}
 & & & edge &  & & & total & back- &  &  & best & sdp & total & back- \\
name & $n$ & $m$ & density & $\alpha$ & & $\alpha$ & time & tracks & & $\alpha$ & discr & time & time & tracks \\ \hline
1dc.64 & 64 & 543 & 0.27 & 10 &  & 10 & 11.44 & 79519 &  & 10 & 0 & 5.08 & 5.09 & 0 \\
1dc.128 & 128 & 1471 & 0.18 & 16 &  & $\geq$ 16 & limit &  &  & 16 & 0 & 49.95 & 49.98 & 0 \\
1dc.256 & 256 & 3839 & 0.12 & 30 &  & $\geq$ 26 & limit &  &  & 30 & 0 & 882.21 & 882.33 & 0 \\
1et.64 & 64 & 264 & 0.13 & 18 &  & 18 & 273.06 & 2312832 &  & 18 & 0 & 1.07 & 1.08 & 0 \\
1et.128 & 128 & 672 & 0.08 & 28 &  & $\geq$ 28 & limit &  &  & $\geq$ 28 & 0 & 11.22 & limit &  \\
1et.256 & 256 & 1664 & 0.05 & 50 &  & $\geq$ 46 & limit &  &  & $\geq$ 50 & 0 & 107.58 & limit &  \\
1tc.64 & 64 & 192 & 0.10 & 20 &  & $\geq$ 20 & limit &  &  & 20 & 0 & 0.78 & 0.79 & 0 \\
1tc.128 & 128 & 512 & 0.06 & 38 &  & $\geq$ 37 & limit &  &  & 38 & 0 & 8.14 & 8.18 & 0 \\
1tc.256 & 256 & 1312 & 0.04 & 63 &  & $\geq$ 58 & limit &  &  & $\geq$ 63 & 4 & 72.75 & limit &  \\
1tc.512 & 512 & 3264 & 0.02 & 110 &  & $\geq$ 100 & limit &  &  & $\geq$ 110 & 2 & 719.56 & limit & \\ 
1zc.128 & 128 & 2240 & 0.28 & 18 &  & $\geq$ 18 & limit &  &  & $\geq$ 18 & 4 & 129.86 & limit &  \\ \hline\hline
\\
\end{tabular}
\normalsize
\end{center}
\end{table}

The results are reported in Table~\ref{tb:sloane}, which follows the
same format as Table~\ref{tb:random}. It shows the same behaviour as
the results on random graphs. Namely, our method always finds better 
solutions than the standard constraint programming solver, in less 
time or within the time limit. This is not surprising, because the 
edge density of these instances are exactly in the region in which 
our method is supposed to behave best (with the exception of 1dc.64 
and 1zc.128), as analyzed in Section~\ref{ssc:properties}. Again, 
our method finds the best solutions at a low discrepancy. Note that 
the instance 1et.64 shows the strength of the semidefinite relaxation 
with respect to standard constraint programming. The difference in 
computation time to prove optimality is huge.

\subsection{Graphs from the DIMACS Benchmarks Set}
Our final experiments are performed on a subset of the DIMACS benchmark 
set for the maximum clique problem \cite{dimacs_clique}. Although our 
method is not intended to be competitive with the best heuristics and 
exact methods for maximum clique problems, it is still interesting to 
see its performance on this standard benchmark set. As pointed out
in Section~\ref{sc:formulations}, we have transformed these maximum 
clique problems to stable set problems on the complement graph.

\begin{table}
\caption{Computational results on graphs from the DIMACS benchmark set
for maximum clique problems, with $n$ vertices and $m$ edges. All times 
are in seconds. The time limit is set to 1000 seconds.}
\label{tb:dimacs}
\begin{center}
\renewcommand{\arraystretch}{1.1}
\setlength{\tabcolsep}{0.09cm}
\tiny
\setlength{\doublerulesep}{\arrayrulewidth} 
\begin{tabular}{lrrcrcrrrcrcrrr} \\ \hline\hline
\multicolumn{5}{c}{\bf instance} & & \multicolumn{3}{c}{\bf CP} & & \multicolumn{5}{c}{\bf SDP + CP} \\ 
\cline{1-5} \cline{7-9} \cline{11-15}
 & & & edge &  & & & total & back- &  &  & best & sdp & total & back- \\
name & $n$ & $m$ & density & $\alpha$ & & $\alpha$ & time & tracks & & $\alpha$ & discr & time & time & tracks \\ \hline
hamming6-2 & 64 & 192 & 0.095 & 32 &  & 32 & 20.22 & 140172 &  & 32 & 0 & 0.68 & 0.69 & 0 \\
hamming6-4 & 64 & 1312 & 0.651 & 4 &  & 4 & 0.28 & 804 &  & 4 & 0 & 27.29 & 28.10 & 706 \\
hamming8-2 & 256 & 1024 & 0.031 & 128 &  & $\geq$ 128 & limit &  &  & 128 & 0 & 45.16 & 45.55 & 0 \\
johnson8-2-4 & 28 & 168 & 0.444 & 4 &  & 4 & 0.05 & 255 &  & 4 & 0 & 0.35 & 0.35 & 0 \\
johnson8-4-4 & 70 & 560 & 0.232 & 14 &  & 14 & 15.05 & 100156 &  & 14 & 0 & 4.82 & 4.83 & 0 \\
johnson16-2-4 & 120 & 1680 & 0.235 & 8 &  & $\geq$ 8 & limit &  &  & 8 & 0 & 43.29 & 43.32 & 0 \\
MANN\_a9 & 45 & 72 & 0.072 & 16 &  & 16 & 162.81 & 1738506 &  & 16 & 1 & 0.17 & 82.46 & 411104 \\
MANN\_a27 & 378 & 702 & 0.010 & 126 &  & $\geq$ 103 & limit &  &  & $\geq$ 125 & 3 & 70.29 & limit &  \\
MANN\_a45 & 1035 & 1980 & 0.004 & 345 &  & $\geq$ 156 & limit &  &  & $\geq$ 338 & 1 & 1047.06 & limit & \\ 
san200\_0.9\_1 & 200 & 1990 & 0.100 & 70 &  & $\geq$ 45 & limit &  &  & 70 & 0 & 170.01 & 170.19 & 0 \\
san200\_0.9\_2 & 200 & 1990 & 0.100 & 60 &  & $\geq$ 36 & limit &  &  & 60 & 0 & 169.35 & 169.51 & 0 \\
san200\_0.9\_3 & 200 & 1990 & 0.100 & 44 &  & $\geq$ 26 & limit &  &  & 44 & 0 & 157.90 & 157.99 & 0 \\
sanr200\_0.9 & 200 & 2037 & 0.102 & 42 &  & $\geq$ 34 & limit &  &  & $\geq$ 41 & 4 & 131.57 & limit &  \\ \hline\hline
\\
\end{tabular}
\normalsize
\end{center}
\end{table}

The results are reported in Table~\ref{tb:dimacs}, which again follows
the same format as Table~\ref{tb:random}. The choice for this 
particular subset of instances is made by the solvability of an 
instance by a semidefinite programming solver in reasonable time 
(again, reasonable means 1000 seconds). For all instances with 
edge density smaller than 0.24, our method outperforms the standard
constraint programming approach. For higher densities however, the 
opposite holds. This is exactly what could be expected from the 
analysis of Section~\ref{ssc:properties}.
A special treatment has been given to instance MANN\_a45. We stopped 
the semidefinite programming solver at the time limit of 1000 seconds, 
and used its intermediate feasible solution as if it were the optimal 
fractional solution. We then proceeded our algorithm for a couple of 
seconds more, to search for a solution up to discrepancy 1.

\begin{table}
\caption{A comparison of different methods on graphs from the DIMACS 
benchmark set for maximum clique problems, with $n$ vertices and $m$ 
edges. All times are in seconds.}
\label{tb:compare}
\begin{center}
\renewcommand{\arraystretch}{1.1}
\setlength{\tabcolsep}{0.12cm}
\tiny
\setlength{\doublerulesep}{\arrayrulewidth} 
\begin{tabular}{lcr c rr c rrr c rrr} \\ \hline\hline
\multicolumn{3}{c}{\bf instance} & & \multicolumn{2}{c}{\bf {\"O}sterg{\aa}rd} & & \multicolumn{3}{c}{\bf R\'egin} & & \multicolumn{3}{c}{\bf CP + SDP} \\
\cline{1-3} \cline{5-6} \cline{8-10} \cline{12-14}
 & edge &  & & & total &  & & total & back- & & & total & back- \\
name & density & $\alpha$ & & $\alpha$ & time & & $\alpha$ & time & tracks & & $\alpha$ & time & tracks \\ \hline
hamming6-2 & 0.095 & 32 &  & 32 & 0.01 &  & 32 & 0.00 & 17 &  & 32 & 0.69 & 0 \\
hamming6-4 & 0.651 & 4 &  & 4 & 0.01 &  & 4 & 0.00 & 42 &  & 4 & 28.10 & 706 \\
hamming8-2 & 0.031 & 128 &  & 128 & 0.04 &  & 128 & 0.00 & 65 &  & 128 & 45.55 & 0 \\
johnson8-2-4 & 0.444 & 4 &  & 16 & 0.01 &  & 4 & 0.00 & 14 &  & 4 & 0.35 & 0 \\
johnson8-4-4 & 0.232 & 14 &  & 14 & 0.01 &  & 14 & 0.00 & 140 &  & 14 & 4.83 & 0 \\
johnson16-2-4 & 0.235 & 8 &  & 8 & 0.27 &  & 8 & 11.40 & 250505 &  & 8 & 43.32 & 0 \\
MANN\_a9 & 0.073 & 16 &  & 16 & 0.01 &  & 16 & 0.00 & 50 &  & 16 & 82.46 & 411104 \\
MANN\_a27 & 0.010 & 126 &  &  & $>$ 10000 &  & 126 & 55.44 & 1258768 &  & $\geq$ 125 & $>$ 1000 &  \\
MANN\_a45 & 0.004 & 345 &  &  & $>$ 10000 &  & $\geq$ 345 & $>$ 43200 &  &  & $\geq$ 338 & $>$ 1000 &  \\
san200\_0.9\_1 & 0.100 & 70 &  & 70 & 0.27 &  & 70 & 3.12 & 1040 &  & 70 & 170.19 & 0 \\
san200\_0.9\_2 & 0.100 & 60 &  & 60 & 4.28 &  & 60 & 7.86 & 6638 &  & 60 & 169.51 & 0 \\
san200\_0.9\_3 & 0.100 & 44 &  &  & $>$ 10000 &  & 44 & 548.10 & 758545 &  & 44 & 157.99 & 0 \\
sanr200\_0.9 & 0.102 & 42 &  &  & $>$ 10000 &  & 42 & 450.24 & 541496 &  & $\geq$ 41 & $>$ 1000 &  \\ \hline\hline
\\
\end{tabular}
\normalsize
\end{center}
\end{table}

In Table~\ref{tb:compare} we compare our method with two methods that
are specialized for maximum clique problems. The first method was 
presented by \"Osterg{\aa}rd~\cite{ostergard}, and follows a 
branch-and-bound approach. The second method is a constraint 
programming approach, using a special constraint for the maximum 
clique problem, with a corresponding propagation algorithm. 
This idea was introduced by Fahle~\cite{fahle_esa02} 
and extended and improved by R\'egin~\cite{regin_cp03}. Since all 
methods are performed on different machines, we need to identify
a time ratio between them. A machine comparison from 
SPEC\footnote{\tt http://www.spec.org/} shows that our times are 
comparable with the times of \"Osterg{\aa}rd. We have multiplied the 
times of R\'egin with 3, following the time comparison made in 
\cite{regin_cp03}. In general, our method is outperformed by the 
other two methods, although there is one instance on which our method 
performs best (san200\_0.9\_3).

\section{Conclusion and Future Work} \label{sc:conclusions}
We have presented a method to use semidefinite relaxations within
constraint programming. The fractional solution values of the 
relaxation serve as an indication for the corresponding constraint
programming variables. Moreover, the solution value of the 
relaxation is used as a bound for the corresponding constraint
programming objective function.

We have implemented our method to find the maximum stable set in a 
graph. Experiments are performed on random weighted and unweighted 
graphs, structured graphs from coding theory, and on a subset of
the DIMACS benchmarks set for maximum clique problems.
Computational results show that constraint programming can greatly 
benefit from semidefinite programming. Indeed, the solution to the
semidefinite relaxations turn out to be very informative. Compared to 
a standard constraint programming approach, our method obtains far 
better results. Specialized algorithms for the maximum clique problem 
however, generally outperform our method.

The current work has investigated the possibility of exploiting 
semidefinite relaxations in constraint programming. Possible future
investigations include the comparison of our method with methods 
that use linear relaxations, for instance branch-and-cut algorithms.
Moreover, one may investigate the effects of strengthening the 
relaxation by adding redundant constraints. For instance, for the 
stable set problem so-called clique-constraints (among others) can 
be added to both the semidefinite and the linear relaxation.
Finally, the proof of optimality may be accelerated similar to a
method presented in~\cite{milano_hoeve02}, by adding 
so-called discrepancy constraints to the semidefinite or the linear 
relaxation, and recompute the solution to the relaxation.

\section*{Acknowledgements}
Many thanks to Michela Milano, Monique Laurent and Sebastian Brand 
for fruitful discussions and helpful comments while writing 
(earlier drafts of) this paper. Also thanks to the anonymous 
referees for useful comments.

\end{document}